\documentclass[english]{article}
\usepackage[T1]{fontenc}
\usepackage[latin9]{inputenc}
\usepackage{geometry}
\usepackage{color}
\usepackage{array}
\usepackage{float}
\usepackage{amstext}
\usepackage{graphicx}

\makeatletter

\providecommand{\tabularnewline}{\\}

\usepackage{epstopdf}
\usepackage{epsfig}

\makeatother

\usepackage{babel}
\begin{document}

\title{Applications of quantum cryptographic switch: Various tasks related
to controlled quantum communication can be performed using Bell states
and permutation of particles}

\author{Kishore Thapliyal and Anirban Pathak}

\maketitle
\begin{center}
Jaypee Institute of Information Technology, A-10, Sector-62, Noida,
UP-201307, India
\par\end{center}
\begin{abstract}
Recently, several aspects of controlled quantum communication (e.g.,
bidirectional controlled state teleportation, controlled quantum secure
direct communication, controlled quantum dialogue, etc.) have been
studied using $n$-qubit ($n\geq3$) entanglement. Specially, a large
number of schemes for bidirectional controlled state teleportation
are proposed using $m$-qubit entanglement ($m\in\{5,6,7\}$). Here,
we propose a set of protocols to illustrate that it is possible to
realize all these tasks related to controlled quantum communication
using only Bell states and permutation of particles (PoP). As the
generation and maintenance of a Bell state is much easier than a multi-partite
entanglement, the proposed strategy has a clear advantage over the
existing proposals. Further, it is shown that all the schemes proposed
here may be viewed as applications of the concept of quantum cryptographic
switch which was recently introduced by some of us. The performances
of the proposed protocols as subjected to the amplitude damping and
phase damping noise on the channels are also discussed.
\end{abstract}

\section{Introduction}

In 1993, Bennett \textit{et al}. \cite{Bennett} introduced the fascinating
idea of quantum teleportation. Since then a large number of modified
teleportation schemes have been proposed. For example, all the existing
schemes of quantum information splitting (QIS) or controlled teleportation
(CT) \cite{Ct,A.Pathak}, quantum secret sharing (QSS) \cite{Hillery},
hierarchical quantum information splitting (HQIS) \cite{hierarchical,Shukla},
remote state preparation \cite{Pati}, etc., can be viewed as modified
teleportation schemes. The absence of any classical analogue and the
potential applications in secure quantum communication and remote
quantum operations \cite{J. A. Vaccaro} made the studies on teleportation
and modified teleportation schemes extremely important. Bennett \textit{et
al}.'s original scheme \cite{Bennett} of quantum teleportation was
a one-way scheme which enables Alice (sender) to transmit an unknown
quantum state (single qubit quantum state) to Bob (receiver) by using
two bits of classical communication and a pre-shared maximally entangled
state. Later on, Huelga \textit{et al}. \cite{J. A. Vaccaro,J. A. Vaccaro-1}
and others generalized the original scheme of Bennett \emph{et al}.
to design the protocols for bidirectional state teleportation (BST).
In a BST scheme, Alice and Bob are allowed to simultaneously transmit
unknown quantum states to each other. Huelga \textit{et al}. also
established that BST can be used to implement nonlocal quantum gates.
This can be visualized clearly, if we consider that Bob teleports
a quantum state $|\psi\rangle$ to Alice, who applies a unitary operator
$U$ on $|\psi\rangle$ and teleports back the state $|\psi^{\prime}\rangle=U|\psi\rangle$
to Bob. Thus, the existence of a scheme for BST essentially implies
the ability to implement a nonlocal quantum gate or a quantum remote
control. This is why BST schemes are very important for both quantum
computation and quantum communication. Further extending the idea
of BST, a good number of protocols for bidirectional controlled state
teleportation (BCST) have been proposed in the recent past \cite{Zha,Zha II,Li,bi-directional-ourpaper,5-qubit-c-qsdc,sixqubit1,sixqubiit2,six-qubit-3,six-qubit-4,7qubit}.
A BCST scheme is a three party scheme, where BST is possible provided
the supervisor/controller (Charlie) permits the other two users (Alice
and Bob) to execute a protocol of BST. A careful study of all the
recently proposed schemes of BCST \cite{Zha,Zha II,Li,5-qubit-c-qsdc,sixqubit1,sixqubiit2,six-qubit-3,six-qubit-4,7qubit}
reveals that different $n$-qubit (with $n\geq5)$ entangled states
are used in these protocols. Thus, at least 5 qubits are used in all
the existing protocols of BCST. Keeping this in mind, in an earlier
work \cite{bi-directional-ourpaper}, we explored the intrinsic symmetry
of the 5-qubit quantum states that were used to propose the protocols
of BCST until then. In Ref. \cite{bi-directional-ourpaper}, we reported
a general structure of the quantum states that can be used for BCST
as follows: 
\begin{equation}
|\psi\rangle=\frac{1}{\sqrt{2}}\left(|\psi_{1}\rangle_{A_{1}B_{1}}|\psi_{2}\rangle_{A_{2}B_{2}}|a\rangle_{C_{1}}\pm|\psi_{3}\rangle_{A_{1}B_{1}}|\psi_{4}\rangle_{A_{2}B_{2}}|b\rangle_{C_{1}}\right),\label{eq:the 5-qubit state}
\end{equation}
where single qubit states $|a\rangle$ and $|b\rangle$ satisfy $\langle a|b\rangle=\delta_{a,b}$,
$|\psi_{i}\rangle\in\left\{ |\psi^{+}\rangle,|\psi^{-}\rangle,|\phi^{+}\rangle,|\phi^{-}\rangle:|\psi_{1}\rangle\neq|\psi_{3}\rangle,|\psi_{2}\rangle\neq|\psi_{4}\rangle\right\} $,
$|\psi^{\pm}\rangle=\frac{|00\rangle\pm|11\rangle}{\sqrt{2}},$ $|\phi^{\pm}\rangle=\frac{|01\rangle\pm|10\rangle}{\sqrt{2}}$,
and the subscripts $A$, $B$ and $C$ indicate the qubits of Alice,
Bob and Charlie, respectively. Charlie prepares the state $|\psi\rangle$
and sends 1st and 3rd (2nd and 4th) qubits to Alice (Bob) and keeps
the last qubit with himself. The condition 
\begin{equation}
|\psi_{1}\rangle\neq|\psi_{3}\rangle,|\psi_{2}\rangle\neq|\psi_{4}\rangle\label{eq:condition}
\end{equation}
ensures that Charlie's qubit is appropriately entangled with the remaining
4 qubits, and this in turn ensures that Alice and Bob are unaware
of the entangled (Bell) states they share until Charlie measures his
qubit using $\{|a\rangle,|b\rangle\}$ basis and discloses the outcome.
Once Charlie discloses the measurement outcome, Alice and Bob know
with certainty which two Bell states they share, and consequently
they can use the conventional teleportation scheme to teleport unknown
quantum states to each other by using classical communication and
the unitary operations described in Table \ref{tab:table1}. It is
important to note that in a successful protocol of BCST, Charlie should
have control over both the directions of communication (i.e., both
Alice to Bob and Bob to Alice transmission channels). We specifically
note this because there exist a few proposals for BCST, where Charlie's
control is limited to one direction only \cite{Li,5-qubit-c-qsdc}.
For example, the five qubit state used in Ref. \cite{5-qubit-c-qsdc}
does not satisfy (\ref{eq:condition}), and consequently it does not
provide a scheme for controlled bidirectional teleportation (cf. Eq.
(3) of Ref. \cite{5-qubit-c-qsdc} where the qubits 13 are separable
from the qubits 245 implying that Alice and Bob can ignore Charlie's
control in one of the directions and teleport using the qubits 13).
Similar shortcoming exists in the BCST protocol proposed by Li \textit{et
al}. \cite{Li}, and the same was noted in our earlier work \cite{bi-directional-ourpaper}.
These two protocols do not qualify as the protocols of BCST. However,
the other recently proposed protocols do not have this limitation.

\begin{table}
\begin{centering}
\begin{tabular}{|c|c|c|c|c|}
\hline 
 & \multicolumn{4}{c|}{Initial state shared by Alice and Bob}\tabularnewline
\hline 
 & $|\psi^{+}\rangle$  & $|\psi^{-}\rangle$  & $|\phi^{+}\rangle$  & $|\phi^{-}\rangle$\tabularnewline
\hline 
SMO  & Receiver & Receiver  & Receiver  & Receiver\tabularnewline
\hline 
00  & $I$  & $Z$  & $X$  & $iY$\tabularnewline
\hline 
01  & $X$  & $iY$  & $I$  & $Z$\tabularnewline
\hline 
10  & $Z$  & $I$  & $iY$  & $X$\tabularnewline
\hline 
11  & $iY$  & $X$ & $Z$  & $I$\tabularnewline
\hline 
\end{tabular}
\par\end{centering}

\caption{\label{tab:table1} Perfect Teleportation. Here SMO stands for sender's
measurement outcome.}
\end{table}

In the above scheme, we require at least 5-qubit entanglement, if
we choose $|a\rangle$ and $|b\rangle$ as single qubit states. One
can of course achieve the same thing using a 6 or more qubit states
by choosing $|a\rangle$ and $|b\rangle$ as multi-qubit states. For
example, in Refs. \cite{sixqubit1,sixqubiit2,six-qubit-3,six-qubit-4}
BCST is reported using 6-qubit entangled states, and in Ref. \cite{7qubit}
BCST is reported using a 7-qubit entangled state. In Ref. \cite{sixqubit1},
it is explicitly shown that the 6-qubit entangled state used in \cite{sixqubit1}
is of the form (\ref{eq:the 5-qubit state}) where $|a\rangle$ and
$|b\rangle$ were chosen as 2-qubit states and the six-qubit entangled
state used in \cite{six-qubit-4} can also be expressed in a form
which can be viewed as an extended form of (\ref{eq:the 5-qubit state})
that follows the same logic%
\footnote{By the same logic we mean that unless Charlie measures his qubits
and discloses the results Alice and Bob do not know which Bell states
they share.%
} (cf. Eqs. (11)-(12) of Ref. \cite{sixqubit1}). It is well-known
that the experimental generation and maintenance of a 6-qubit or 7-qubit
state is extremely difficult at the moment. This fact motivated us
to look into the possibility of achieving BCST solely using Bell states
(minimal quantum resource). Further, a few protocols for other controlled
quantum communication tasks have been proposed in the recent past
using multi-qubit states \cite{5-qubit-c-qsdc,C-QD1,C-QD2,CQSDC-Hassanpour}.
Specifically, in Refs. \cite{C-QD1,C-QD2} 3-qubit GHZ states were
used to design protocols for controlled quantum dialogue (CQD), and
in Ref. \cite{5-qubit-c-qsdc} a 5-qubit quantum state was used to
design a protocol of controlled quantum secure direct communication
(CQSDC). In what follows, we will establish that it is possible to
design protocols of CQD and CQSDC solely using Bell states.

In our earlier work on BCST \cite{bi-directional-ourpaper}, we just
mentioned that BCST may be viewed as a quantum cryptographic switch
\cite{switch}, a concept introduced by us in a different context.
Here, we elaborate the link and show that using the idea of quantum
cryptographic switch, it is possible to construct schemes for BCST
and other controlled quantum communication tasks solely using Bell
states. Thus, to implement BCST, one does not require multi-partite
entangled states as described in the recent papers \cite{Zha,Zha II,Li,bi-directional-ourpaper,5-qubit-c-qsdc,sixqubit1,sixqubiit2,six-qubit-3,six-qubit-4,7qubit},
rather one can construct protocols of BCST solely using Bell states
and a technique known as permutation of particles (PoP), which was
first introduced in 2003 by Deng and Long \cite{PoP}. Since the pioneering
work of Deng and Long, this interesting technique has been used in
many protocols (\cite{With chitra IJQI,With chitra-ijtp,With Anindita-pla,With preeti}
and references therein). Before we describe our Bell-state-based protocols,
it would be apt to briefly introduce the concept of quantum cryptographic
switch, as it plays a crucial role in the construction of the new
protocols described in the present paper. In a quantum scenario, a
cryptographic switch describes a situation in which a controller/supervisor
(Charlie) can control to a continuously varying degree the amount
of information received by a semi-honest receiver (Bob), after a semi-honest
sender (Alice) sends her information through a quantum channel. To
understand the function of quantum cryptographic switch, let us consider
that Charlie transmits a Bell state to Alice and Bob, but does not
disclose which Bell state it is. Subsequently, Alice uses dense coding
to transmit two bits of classical information to Bob. However, Bob
can perfectly disclose the information encoded by Alice iff Charlie
discloses the 2-bit of information corresponding to his choice of
the Bell state. Charlie can continually vary the amount of disclosed
information, and thus he can continuously alter the amount of information
recovered by Bob. Further, we would like to note that in all the BCST
protocols described until now, Alice and Bob need to be semi-honest
as otherwise, they can completely ignore the supervisor Charlie and
prepare and share entanglement by themselves. Such possibilities can
be circumvented by specifically mentioning Alice and Bob as semi-honest
(a user who follows the protocol but tries to cheat the controller).
Unfortunately, this requirement was not specified in the existing
protocols.

The remaining part of the present paper is organized as follows. In
Section \ref{sec:Bidirectional-teleportation-using}, we describe
a Bell-state-based protocol of BCST. In Section \ref{sec:Other-protocols-of},
we describe a set of Bell-state-based protocols for controlled quantum
communication. Specifically, we explicitly describe a protocol of
CQD and discuss how to obtain similar protocols for CQSDC, controlled
quantum key agreement (CQKA), controlled quantum key distribution
(CQKD), etc. In Section \ref{sec:Effect-of-noise}, we study the effect
of amplitude damping and phase damping noise on the Bell-state-based
protocol for BCST. Finally, the paper is concluded in Section \ref{sec:Conclusions}.

\section{Bidirectional controlled teleportation using Bell states\label{sec:Bidirectional-teleportation-using}}

If Alice and Bob are semi-honest then our protocol of BCST using Bell
states works as follows
\begin{enumerate}
\item Charlie prepares $2n$ Bell states with $n\geq1$ (each of them is
randomly prepared in one of the Bell states%
\footnote{We can assume that Charlie has a quantum random number generator,
and he has generated a large sequence of 0 and 1 through it. He uses
the outcomes of the random number generator to decide which Bell state
is to be prepared. For example, we may consider that if the first
two bit values obtained from the random number generator are 00,01,10,
and 11 then he prepares $|\psi^{+}\rangle,|\psi^{-}\rangle,|\phi^{+}\rangle,$
and $|\phi^{-}\rangle,$ respectively. Here, $|\psi^{\pm}\rangle=\frac{1}{\sqrt{2}}\left(|00\rangle\pm|11\rangle\right)$
and $|\phi^{\pm}\rangle=\frac{1}{\sqrt{2}}\left(|01\rangle\pm|10\rangle\right)$
are Bell states.%
}). He uses the Bell states to prepare 4 ordered sequences as follows: 

\begin{enumerate}
\item A sequence with all the first qubits of the first $n$ Bell states:
$P_{A_{1}}=\left[p_{1}\left(t_{A}\right),p_{2}\left(t_{A}\right),...,p_{n}\left(t_{A}\right)\right]$, 
\item A sequence with all the first qubits of the last $n$ Bell states:
$P_{A_{2}}=\left[p_{n+1}\left(t_{A}\right),p_{n+2}\left(t_{A}\right),...,p_{2n}\left(t_{A}\right)\right]$,
\item A sequence with all the second qubits of the first $n$ Bell states:
$P_{B_{1}}=[p_{1}(t_{B}),p_{2}(t_{B}),...,p_{n}(t_{B})]$,
\item A sequence with all the second qubits of the last $n$ Bell states:
$P_{B_{2}}=[p_{n+1}(t_{B}),p_{n+2}(t_{B}),...,p_{2n}(t_{B})]$.
\end{enumerate}

where the subscript $1,2,\cdots,2n$ denote the order of a particle
pair $p_{i}=\{t_{A}^{i},t_{B}^{i}\},$ which is in the Bell state. 

\item Charlie randomizes the sequences of the second qubits, i.e., he applies
$n$-qubit permutation operators $\Pi_{n_{1}}$ and $\Pi_{n_{2}}$
on $P_{B_{1}}$ and $P_{B_{2}}$ to create two new sequences as $P_{B_{i}}^{\prime}=\Pi_{n_{i}}P_{B_{i}}$
with $i\in\left\{ 1,2\right\} $ and sends these sequences to Bob.
The actual order is known to Charlie only. Charlie also sends $P_{A_{1}}$
and $P_{A_{2}}$ to Alice. \\
It is predeclared that the first (last) $n$ Bell states prepared
by Charlie are to be used for Alice to Bob (Bob to Alice) teleportation. 
\item After receiving the qubits from Charlie, Alice (Bob) understands that
she (he) can now teleport her (his) unknown qubits $|\psi_{A_{j}}\rangle=\alpha_{A_{j}}|0\rangle+\beta_{A_{j}}|1\rangle:\,|\alpha_{A_{j}}|^{2}+|\beta_{A_{j}}|^{2}=1$
($|\psi_{B_{j}}\rangle=\alpha_{B_{j}}|0\rangle+\beta_{B_{j}}|1\rangle:\,|\alpha_{B_{j}}|^{2}+|\beta_{B_{j}}|^{2}=1$)
to Bob (Alice) using standard teleportation scheme, i.e., by entangling
her (his) unknown qubit $|\psi_{A_{j}}\rangle$ ($|\psi_{B_{j}}\rangle$)
with $p_{j}^{\prime}(t_{A})$ ($p_{n+j}^{\prime}(t_{B})$) and subsequently
measuring both the qubits in computational basis and communicating
the result to Bob (Alice). \\
Since the sequence with Alice and Bob are different, even if Alice
or Bob obtain the access of both $P_{A_{i}}$ and $P_{B_{i}}^{\prime}$,
they will not be able to find out the Bell states prepared by Charlie.
Thus, any kind of collusion between Alice and Bob would not help Alice
and Bob to circumvent the control of Charlie as long as they are semi-honest.
Specifically, even after Alice (Bob) communicates the outcome of her
(his) measurement in computational basis to Bob (Alice), he (she)
will not be able to reproduce the unknown state sent by Alice (Bob)
as he (she) neither knows which Bell state was prepared by Charlie
nor knows which qubit was entangled with which qubit. Interestingly,
the existing protocols of BCST are not protected under such a collusion
between Alice and Bob as in all the protocols that uses quantum states
of the form (\ref{eq:the 5-qubit state}) Alice (Bob) can send her
qubits to Bob (Alice) so that Bob (Alice) can perform an appropriate
Bell measurement to know the nature of Bell state and return the particles
to Alice (Bob).
\item When Charlie plans to allow Bob (Alice) to reconstruct the unknown
quantum state teleported by Alice (Bob), then Charlie discloses the
Bell state which he had prepared and the exact sequence $\Pi_{n_{1}}$
$\left(\Pi_{n_{2}}\right)$. 
\item Since the initial Bell states and the exact sequence are known, Bob
(Alice) now applies appropriate unitary operations on his (her) qubits
as described in Table \ref{tab:table1} and obtains the unknown quantum
states sent by Alice (Bob). 
\end{enumerate}
This protocol has certain advantages over the existing protocols of
BCST. Firstly, it requires only 2-qubit entanglement which is much
easier to produce and maintain. Secondly, the control of Charlie on
Alice and Bob is much more compared to that in the existing protocols.
To be precise, using $\Pi_{n_{1}}$ and $\Pi_{n_{2}}$ Charlie can
separately control Alice to Bob and Bob to Alice teleportation channels.
For example, if he discloses $\Pi_{n_{1}}$ and the nature of first
$n$ Bell states prepared by him but keeps $\Pi_{n_{2}}$ and the
nature of the last $n$ Bell states secret, then Bob will be able
to reproduce the state teleported by Alice, but Alice will not be
able to obtain the state teleported by Bob. Such directional control
was missing in earlier protocols of BCST. For example, in all the
protocols that use a state of the form (\ref{eq:the 5-qubit state}),
when Charlie announces his measurement outcome in $\left\{ |a\rangle,|b\rangle\right\} $
basis, then both Alice and Bob will know the entangled states shared
by them and will be able to reconstruct the state teleported by their
partners. Further, in all the previous protocols Alice (Bob) always
prepares the state sent by Bob (Alice) with perfect fidelity if the
channel is perfect. Charlie, has no control over the fidelity with
which Bob (Alice) can reconstruct the state teleported by Alice (Bob).
However, in the present protocol, which is a variant of quantum cryptographic
switch, maximum fidelity of the state prepared by Alice and Bob can
be controlled by Charlie. This is so because in the present protocol
Charlie has the freedom to decide whether he will precisely disclose
which Bell state he had prepared (a 2-bit information) in a particular
position. For example, Charlie may choose to inform Bob that the first
Bell state prepared by him was such that the parity-0 Bell states
are twice more likely than the parity-1 states and that Bell states
of equal parity are equally likely. This corresponds to a probability
distribution of $\left(\frac{1}{3},\frac{1}{3},\frac{1}{6},\frac{1}{6}\right)$,
i.e., an entropy of about 1.92 bits, implying that Charlie reveals
${\rm c}=0.08$ bits. Consequence of such an incomplete disclosure
is that Bob will not be able to uniquely decide which unitary operations
to be applied, and consequently the fidelity of the state reconstructed
by Bob will reduce. In brief, the BCST protocol proposed here requires
lesser quantum resources and provides better control. Further, it
may be modified to a multi-controlled BST (MCBST) scheme where first
$n$ Bell states are prepared by Charlie 1 and the last $n$ Bell
states are prepared by Charlie 2. Thus, the BST scheme will not work
unless both of them disclose the exact sequences and what Bell states
were prepared by them. However, this simple idea of MCBST has a limitation
that one Charlie controls one channel (say Alice to Bob) and the other
Charlie controls the other Channel (say Bob to Alice). A more sophisticated
idea of MCBST will be separately discussed in a future work.

\section{Other protocols of controlled quantum communication using Bell states
\label{sec:Other-protocols-of}}

A large number of entangled-state-based quantum communication protocols
for different communication tasks have been proposed in the recent
past. For example, protocols for quantum dialogue (QD) \cite{ba-an,qd},
quantum key distribution (QKD) \cite{bb84,ekert}, quantum secure
direct communication (QSDC), deterministic secure quantum communication
(DSQC) \cite{With chitra-ijtp,With Anindita-pla,With preeti,ping-pong,CL,DLL}
have been proposed. Using the PoP-based trick described above, one
can obtain the controlled versions of all these two-party (mostly
bidirectional) quantum communication schemes just by using Bell states.
In what follows, we briefly describe how to modify the existing two-party
protocols of quantum communication to obtain the corresponding controlled
three-party protocols of quantum communication. To begin with, we
describe a CQD protocol of Ba An type in the following subsection
and subsequently describe how to transform the proposed protocol of
CQD into a Ping-Pong (PP)-type protocol of CQSDC and a protocol of
CQKA.

\subsection{Controlled quantum dialogue protocol of Ba An type}

Let us first describe the Ba An's original two-party scheme of QD
in which both Alice and Bob can simultaneously communicate. This simple
scheme can be described in the following steps \cite{ba-an}: 
\begin{description}
\item [{Step~1}] Bob prepares $|\phi^{+}\rangle^{\otimes n}:\,|\phi^{+}\rangle=\frac{|01\rangle+|10\rangle}{\sqrt{2}}$
and keeps the first qubit of each Bell state ($|\phi^{+}\rangle)$
with himself as home qubit and encodes his secret message by using
the following rule: He encodes $00,01,10$ and $11$ by applying unitary
operations $I,\, X,\, iY$ and $Z$, respectively on the second qubit. 
\item [{Step~2}] Bob sends a sequence of the second qubits (travel qubits)
to Alice and confirms that Alice has received the qubits. 
\item [{Step~3}] Alice encodes her secret message by using the same rule
and returns the travel qubits to Bob. After receiving the encoded
travel qubits Bob appropriately combines them with the sequence of
home qubits and performs Bell measurements on the partner particles
that were initially prepared in $|\phi^{+}\rangle$. 
\item [{Step~4}] Alice discloses whether it was a run in message mode
(MM) or in control mode (CM). In MM, Bob decodes Alice's bits and
announces the outcome of Bell measurement performed by him. Alice
uses that result to decode the bits encoded by Bob. In CM, Alice announces
her encoding value and Bob uses that for eavesdropping check. 
\end{description}
It is straightforward to convert this two-party protocol into an equivalent
three-party protocol, where Charlie is the controller and Alice and
Bob are semi-honest users who want to execute a scheme of QD. The
modified protocol with clear measures of security works as follows:
\begin{description}
\item [{Step~1}] Charlie prepares $n$ copies of a Bell state $|\phi^{+}\rangle=\frac{|01\rangle+|10\rangle}{\sqrt{2}}$.
He prepares two sequences: the first sequence $P_{B_{1}}$ is prepared
with all the first qubits of Bell pairs and the second sequence $P_{B_{2}}$
is prepared with all the second qubits. 
\item [{Step~2}] Charlie applies $n$-qubit permutation operators $\Pi_{n}$
on $P_{B_{1}}$ to create a new sequence $P_{B_{1}}^{\prime}=\Pi_{n}P_{B_{1}}$
and sends both $P_{B_{1}}^{\prime}$and $P_{B_{2}}$ to Bob.
\item [{Step~3}] Bob uses the qubits of $P_{B_{1}}^{\prime}$ ($P_{B_{2}}$)
as home (travel) qubits. He encodes his secret message $00,01,10$
and $11$ by applying unitary operations $U_{0},U_{1},U_{2}$ and
$U_{3}$ respectively on the second qubit (i.e., on the qubits of
sequence $P_{B_{2}})$. Without loss of generality, we may assume
that $U_{0}=I,\, U_{1}=\sigma_{x}=X,\, U_{2}=i\sigma_{y}=iY$ and
$U_{3}=\sigma_{z}=Z,$ where $\sigma_{i}$ are Pauli matrices. Further,
we assume that after the encoding operation the sequence $P_{B_{2}}$
transforms to $Q_{B_{2}}.$
\item [{Step~4}] Bob first prepares $n$ decoy qubits in a random sequence
of $\{|0\rangle,|1\rangle,|+\rangle,|-\rangle\}$, i.e., the decoy
qubit state is $\otimes_{j=1}^{n}|P_{j}\rangle,\,|P_{j}\rangle\in\{|0\rangle,|1\rangle,|+\rangle,|-\rangle\}.$
Bob then randomly inserts the decoy qubits in $Q_{B_{2}}$ to obtain
an enlarged new sequence $R_{B_{2}}$ and sends that to Alice and
confirms that Alice has received the entire sequence. 
\item [{Step~5}] Bob discloses the positions of decoy qubits, and applies
BB84 subroutine%
\footnote{BB84 subroutine \cite{my-book} means eavesdropping is checked by
following a procedure similar to that adopted in the original BB84
protocol. Specifically, BB84 subroutine implies that Alice (Bob) randomly
selects half of the qubits received by her (him) to form a verification
string. She (He) measures the verification qubits randomly in $\left\{ |0\rangle,|1\rangle\right\} $
or $\left\{ |+\rangle,|-\rangle\right\} $ basis and announces the
measurement outcome, the position of that qubit in the string and
the basis used for the particular measurement. Bob (Alice) also measures
the corresponding qubit using the same basis (if needed) and compares
his (her) result with the announced result of Alice (Bob) to detect
eavesdropping.%
} in collaboration with Alice and thus computes the error rate. If
the error rate exceeds the tolerable limit, then Alice and Bob abort
this communication and repeat the procedure from the beginning. Otherwise,
they go on to the next step. \\
All the intercept-resend attacks are detected in this step. Any attack
by Eve will not provide her any meaningful information about the encoding
operation executed by Bob as Eve's access to the Bell state is limited
to a single qubit.
\item [{Step~6}] Alice encodes her secret message by using the same set
of encoding operations as was used by Bob and subsequently randomly
inserts a set of $n$ decoy qubits in her sequence and returns the
new sequence $R_{B_{3}}$ obtained by this method to Bob. 
\item [{Step~7}] After Bob confirms that he has received $R_{B_{3}},$
Alice discloses the positions of the decoy qubits, and Alice and Bob
follow Step 5 to check eavesdropping. If no eavesdropping is found
they move to the next step. 
\item [{Step~8}] Charlie announces the exact sequence of $P_{B_{1}}$.
\item [{Step~9}] Bob uses the information obtained from Charlie to create
$n$ Bell pairs and performs Bell measurements on them. Subsequently,
he announces the outcomes of his Bell measurements. As Bob knows the
initial Bell state, final Bell state and his own encoding operation
he can decode Alice's bits. Similarly, Alice uses the results of
Bell measurements announced by Bob, knowledge of the initial state
and her own encoding operation to decode Bob's bits. 
\end{description}
Note that unless Charlie discloses the exact sequence, Alice and Bob,
who are assumed to be semi-honest, cannot decode each other's information.
Thus, we have a CQD protocol. Now, it is easy to visualize that we
can restrict the protocol to the extent that only Alice communicates
a message (i.e., Bob neither encodes anything nor announces the outcomes
of his Bell measurements) then the above protocol will reduce to a\textcolor{red}{{}
}PP-type protocol of CQSDC in which Alice can communicate a secure
message to Bob. In the original PP protocol \cite{ping-pong}, only
2 encoding operations were used, and subsequently its efficiency was
increased in Cai Li (CL) protocol \cite{CL} by including all 4 encoding
operations that are used here. Thus, the above protocol can actually
provide us an efficient controlled PP protocol, or equivalently a
controlled CL protocol. Further, we know that all QSDC protocols can
be reduced to QKD protocols as Alice may choose to communicate a random
sequence instead of a meaningful message. In that case, the QSDC protocol
obtained by modifying the CQD protocol will reduce to a protocol of
CQKD. Interestingly,\textcolor{red}{{} }the above protocol can also
be transformed to a protocol of CQKA. Where both Alice and Bob contributes
equally to the final key. It is easy to visualize if we consider that
Alice and Bob possess two random strings of keys (say $k_{A}$ and
$k_{B})$ and they use the CQD protocol described above to communicate
these keys to each other and decide that for all future communications
they will use $k=\mbox{\ensuremath{k_{A}\oplus k_{B}}}$ as the shared
key. This relation between QKA and QD is elaborately discussed in
our earlier work \cite{qka}. Before we finish this section, we would
like to note that in the protocol proposed here, we have used the
BB84 subroutine for eavesdropping check, where decoy qubits were prepared
and measured randomly in $X$-basis and $Z$-basis. One can replace
this by a GV-type subroutine described in \cite{With chitra IJQI},
where decoy qubits are prepared as Bell states and thus can perform
all the tasks described above (i.e., CQD, CQSDC, CQKA, CQKD, etc.)
using orthogonal states alone. This is interesting as not many orthogonal
state based protocols of quantum communication are proposed until
now.

We have already described a few protocols of controlled quantum communication
solely using Bell states and PoP technique. The idea is novel as so
far usually $n$-partite ($n\geq3)$ entanglement is used to achieve
controlled quantum communication protocols. Further, in the present
paper we have restricted ourselves to a group of protocols of quantum
communication and showed that they may be modified to obtain corresponding
controlled protocols. Same strategy may be used to generalize other
entangled-state-based protocols of quantum communication, too. For
example, CL protocol \cite{CL}, DLL protocol \cite{DLL}, etc., can
be easily modified to obtain their controlled versions.

\section{Effect of noise \label{sec:Effect-of-noise}}

In this section, we will investigate the effect of noise on the BCST
schemes and the other schemes proposed by us. To begin with, let us
note a few important things: (i) As the qubits to be teleported are
never available in the channel, we can assume that the channel noise
does not have any effect on them. Thus, it would be sufficient to
study the effect of noise on the qubits that travel through the channel
(travel qubits). (ii) Following the same logic, we can assume that
Charlie's qubit is independent of noise. 

As Charlie's measurement reduces the 5-qubit quantum state used earlier
by us in Ref. \cite{bi-directional-ourpaper} to a product of two
Bell states, and only these qubits are exposed to noise, we can easily
understand that the effect of noise on the 5-qubit state based BCST
would be the same as that in the Bell-state-based BCST protocol proposed
here. Keeping this in mind, here we study the effects of different
kinds of noise channels on the Bell-state-based scheme for BCST. The
results obtained here are also applicable to the 5-qubit state based
BCST scheme under the above assumptions.

It is well known that the amplitude-damping noise model is described
by the following set of Kraus operators \cite{nielsen}: 
\begin{equation}
E_{0}^{A}=\left[\begin{array}{cc}
1 & 0\\
0 & \sqrt{1-\eta_{A}}
\end{array}\right],\,\,\,\,\,\,\,\,\,\,\,\,\,\,\, E_{1}^{A}=\left[\begin{array}{cc}
0 & \sqrt{\eta_{A}}\\
0 & 0
\end{array}\right],\label{eq:Krauss-amp-damping}
\end{equation}
where the decoherence rate $\eta_{A}$ ($0\leq\eta_{A}\leq1)$ describes
the probability of error due to amplitude-damping noisy environment
when a travel qubit passes through it. Similarly, the following set
of Kraus operators describes the phase-damping channel \cite{nielsen}:
\begin{equation}
E_{0}^{P}=\sqrt{1-\eta_{P}}\left[\begin{array}{cc}
1 & 0\\
0 & 1
\end{array}\right],\,\,\,\,\,\,\,\,\,\,\,\,\,\,\, E_{1}^{P}=\sqrt{\eta_{P}}\left[\begin{array}{cc}
1 & 0\\
0 & 0
\end{array}\right],\,\,\,\,\,\,\,\,\,\,\,\,\,\,\, E_{2}^{P}=\sqrt{\eta_{P}}\left[\begin{array}{cc}
0 & 0\\
0 & 1
\end{array}\right],\label{eq:Krauss-phase-damping}
\end{equation}
where $\eta_{P}$ ($0\leq\eta_{P}\leq1)$ is the decoherence rate
for the phase-damping noise. In what follows, we assume that Alice
(Sender1) and Bob (Sender2) wish to teleport qubits $|\zeta_{1}\rangle{}_{S_{1}^{\prime}}\equiv$$a_{1}|0\rangle+b_{1}\exp(i\phi_{1})|1\rangle,$
and $|\zeta_{2}\rangle{}_{S_{2}^{\prime}}\equiv a_{2}|0\rangle+b_{2}\exp(i\phi_{2})|1\rangle$
to Bob (Receiver1) and Alice (Receiver2), respectively through a noisy
channel. The channel is either phase damping or amplitude damping
channel. These assumptions help us to study the effect of these two
types of noise channels independently.

\subsection{Effect of noise on the protocols of CBST }

Consider that Charlie has distributed two Bell states to Alice and
Bob as $|\psi\rangle_{S_{1}R_{1}S_{2}R_{2}}=|\psi_{1}\rangle_{S_{1}R_{1}}|\psi_{2}\rangle_{S_{2}R_{2}}$
with $|\psi_{i}\rangle\in\left\{ |\psi^{+}\rangle,|\psi^{-}\rangle,|\phi^{+}\rangle,|\phi^{-}\rangle\right\} $.
Now, Sender1 and Sender2 wish to teleport qubits $|\zeta_{1}\rangle{}_{S_{1}^{\prime}}$
and $|\zeta_{2}\rangle{}_{S_{2}^{\prime}}$ to Receiver1 and Receiver2,
respectively. Thus, the combined state of the system is $|\psi^{\prime}\rangle_{S_{1}R_{1}S_{2}R_{2}S_{1}^{\prime}S_{2}^{\prime}}=|\psi_{1}\rangle_{S_{1}R_{1}}\otimes|\psi_{2}\rangle_{S_{2}R_{2}}\otimes|\zeta_{1}\rangle_{S_{1}^{\prime}}\otimes|\zeta_{2}\rangle_{S_{2}^{\prime}}$
and the corresponding density matrix (after suitable rearrangement
of the particles) is 
\[
\rho=|\psi\rangle_{S_{1}S_{1}^{\prime}R_{1}S_{2}S_{2}^{\prime}R_{2}}{}_{S_{1}S_{1}^{\prime}R_{1}S_{2}S_{2}^{\prime}R_{2}}\langle\psi|.
\]
Now, the noisy environment described by (\ref{eq:Krauss-amp-damping})
or (\ref{eq:Krauss-phase-damping}) transforms the density operator
$\rho$ as 
\begin{equation}
\rho_{k}=\sum_{i,j}E_{i,S_{1}}^{k}\otimes I_{2,S_{1}^{\prime}}\otimes E_{j,R_{1}}^{k}\otimes E_{j,S_{2}}^{k}\otimes I_{2,S_{2}^{\prime}}\otimes E_{i,R_{2}}^{k}\rho\left(E_{i,S_{1}}^{k}\otimes I_{2,S_{1}^{\prime}}\otimes E_{j,R_{1}}^{k}\otimes E_{j,S_{2}}^{k}\otimes I_{2,S_{2}^{\prime}}\otimes E_{i,R_{2}}^{k}\right)^{\dagger}.\label{eq:noise-effected-density-matrix-1}
\end{equation}
For simplicity, we assume that both the qubits sent to Alice (i.e.,
$S_{1}$ and $R_{2}$ qubits) are affected by the same Kraus operator
(same noise) and similarly, the qubits $R_{1}$ and $S_{2}$ sent
to Bob are also affected by the same Kraus operator.\textcolor{red}{{}
}This assumption is justified as Charlie to Alice communication is
done by a quantum channel and Charlie to Bob communication is done
by another quantum channel. In accordance with the Bell-state based
CBST scheme described above, $S_{1}$ and $S_{1}^{\prime}$ qubits
are measured by Alice using computational basis, whereas $S_{2}$
and $S_{2}^{\prime}$ qubits are measured by Bob using the same basis.
Here, we also assume that the measurements of both Alice and Bob yield
$|00\rangle$. To selectively choose these outcomes, we have to apply
the unitary operator 
\[
U=\left(|00\rangle_{S_{1}S_{1}^{\prime}S_{1}S_{1}^{\prime}}\langle00|\right)\otimes I_{2,R_{1}}\otimes\left(|00\rangle_{S_{2}S_{2}^{\prime}S_{2}S_{2}^{\prime}}\langle00|\right)\otimes I_{2,R_{2}}
\]
on $\rho_{k}$ and that would yield an unnormalized quantum state
\[
\rho_{k_{1}}=U\rho_{k}U^{\dagger},
\]
which can be normalized to 
\[
\rho_{k_{2}}=\frac{\rho_{k_{1}}}{{\rm Tr}\left(\rho_{k_{1}}\right)}.
\]
Now, the combined state of the qubits of Receivers 1 and 2 (i.e.,
the state of the qubits $R_{1}$ and $R_{2}$) in a noisy environment
or $\rho_{k_{3}}$ can be obtained from $\rho_{k_{2}}$ by tracing
out the qubits that are already measured. Specifically, 
\[
\rho_{k_{3}}={\rm Tr}{}_{S_{1}S_{1}^{\prime}S_{2}S_{2}^{\prime}}\left(\rho_{k_{2}}\right).
\]
Depending upon the initial Bell states and the measurement outcomes
of the senders, the receiver(s) may have to apply appropriate Pauli
operators on $\rho_{k_{3}}$ to obtain the final quantum state $\rho_{k,{\rm out}}$,
where the index $k$ denotes a specific noise model. In the present
case, we have already assumed that the outcomes of measurements of
both the senders are $|00\rangle$, if we further assume that the
initial state prepared by Charlie as $|\psi^{+}\rangle^{\otimes2},$
the receivers would not require to apply any unitary operator (in
other words, the receivers need to apply the identity operators only).
We have already assumed that Alice (Sender1) and Bob (Sender2) wish
to teleport qubits $|\zeta_{1}\rangle{}_{S_{1}^{\prime}}=a_{1}|0\rangle+b_{1}\exp(i\phi_{1})|1\rangle$
and $|\zeta_{2}\rangle{}_{S_{2}^{\prime}}=a_{2}|0\rangle+b_{2}\exp(i\phi_{2})|1\rangle$
at the side of Bob (Receiver1) and Alice (Receiver2), respectively.
Thus, in the absence of noise, the expected final state is a product
state 
\[
|T\rangle_{R_{1}R_{2}}=\left(a_{1}|0\rangle+b_{1}\exp(i\phi_{1})|1\rangle\right)\otimes\left(a_{2}|0\rangle+b_{2}\exp(i\phi_{2})|1\rangle\right),
\]
where Alice (Receiver2) will have qubit $a_{2}|0\rangle+b_{2}\exp(i\phi_{2})|1\rangle$
in her possession, and Bob (Receiver1) will have $a_{1}|0\rangle+b_{1}\exp(i\phi_{1})|1\rangle$.
As $a_{i}$ and $b_{i}$ are real, and $a_{i}^{2}+b_{i}^{2}=1,$ we
can assume that $a_{i}=\sin\theta_{i}$ and $b_{i}=\cos\theta_{i}$
with $i\in\{1,2\}$. Thus, 
\[
|T\rangle_{R_{1}R_{2}}=\text{sin}\theta_{1}\text{sin}\theta_{2}|00\rangle+\text{cos}\text{\ensuremath{\theta_{2}}}\text{sin}\ensuremath{\theta_{1}}\exp\left(i\phi_{2}\right)|01\rangle+\cos\text{\ensuremath{\theta_{1}}}\text{sin}\ensuremath{\theta_{2}}\exp\left(i\phi_{1}\right)|10\rangle+\text{cos}\text{\ensuremath{\theta_{1}}}\text{cos}\theta_{2}\exp\left(i\left(\phi_{1}+\phi_{2}\right)\right)|11\rangle.
\]
The effect of noise can now be investigated by comparing the quantum
state $\rho_{k{\rm ,out}}$ produced in the noisy environment with
the state $|T\rangle_{R_{1}R_{2}}$ using fidelity 
\begin{equation}
F=\langle T|\rho_{k,{\rm ou}t}|T\rangle,\label{eq:fidelity}
\end{equation}
which is the square of the conventional definition of fidelity%
\footnote{Usually fidelity $F(\sigma,\rho)$ of two quantum states $\rho$ and
$\sigma$ is defined as $F(\sigma,\rho)=Tr\sqrt{\sigma^{\frac{1}{2}}\rho\sigma^{\frac{1}{2}}}.$
However, in the present work, we have used (\ref{eq:fidelity}) as
the definition of fidelity. %
}.

As we have assumed that our Bell-state-based protocol for CBST starts
with the initial state 

\begin{equation}
|\psi\rangle=|\psi^{+}\rangle_{S_{1}R_{1}}|\psi^{+}\rangle_{S_{2}R_{2}},\label{eq:choosen qstate-for-PCBRSP}
\end{equation}
and the outcomes of the measurements of both the receivers are $|00\rangle$,
the above described method of obtaining $\rho_{k,{\rm out}}$ yields

\begin{equation}
\begin{array}{lcl}
\rho_{A,{\rm out}} & = & N_{A}\left(\begin{array}{cccc}
\frac{\left(1+\eta_{A}\right)^{4}\text{cos}^{2}\text{\ensuremath{\theta_{1}}}\text{cos}^{2}\text{\ensuremath{\theta_{2}}}}{(1-\eta_{A})^{4}} & \frac{\text{cos}^{2}\text{\ensuremath{\theta_{1}}}\text{sin}2\text{\ensuremath{\theta_{2}}}\exp(-\text{\ensuremath{i\phi_{2}})}}{2(1-\eta_{A})^{3}} & \frac{\text{sin}2\text{\ensuremath{\theta_{1}}}\text{cos}^{2}\text{\ensuremath{\theta_{2}}}\exp(-\text{\ensuremath{i\phi_{1}})}}{2(1-\eta_{A})^{3}} & \frac{\sin2\text{\ensuremath{\theta_{1}}}\text{sin}2\theta_{2}\exp(-i\phi_{12})}{4(1-\eta_{A})^{2}}\\
\frac{\text{cos}^{2}\text{\ensuremath{\theta_{1}}}\text{sin}2\text{\ensuremath{\theta_{2}}}\exp(\text{\ensuremath{i\phi_{2}})}}{2(1-\eta_{A})^{3}} & \frac{\left\{ \text{cos}^{2}\text{\ensuremath{\theta_{1}}}\text{sin}^{2}\text{\ensuremath{\theta_{2}}}+\eta_{A}^{2}\text{sin}^{2}\text{\ensuremath{\theta_{1}}}\text{cos}^{2}\text{\ensuremath{\theta_{2}}}\right\} }{(1-\eta_{A})^{2}} & \frac{\text{sin}2\text{\ensuremath{\theta_{1}}}\text{sin}2\text{\ensuremath{\theta_{2}}}\exp(-i\Delta\phi)}{4(1-\eta_{A})^{2}} & \frac{\text{sin}2\text{\ensuremath{\theta_{1}}}\text{sin}^{2}\text{\ensuremath{\theta_{2}}}\exp(-\text{\ensuremath{i\phi_{1}})}}{2(1-\eta_{A})}\\
\frac{\text{sin}2\text{\ensuremath{\theta_{1}}}\text{cos}^{2}\text{\ensuremath{\theta_{2}}}\exp(\text{\ensuremath{i\phi_{1}})}}{2(1-\eta_{A})^{3}} & \frac{\text{sin}2\text{\ensuremath{\theta_{1}}}\text{sin}2\text{\ensuremath{\theta_{2}}}\exp(i\Delta\phi)}{4(1-\eta_{A})^{2}} & \frac{\left\{ \eta_{A}^{2}\text{cos}^{2}\text{\ensuremath{\theta_{1}}}\text{sin}^{2}\text{\ensuremath{\theta_{2}}}+\text{sin}^{2}\text{\ensuremath{\theta_{1}}}\text{cos}^{2}\text{\ensuremath{\theta_{2}}}\right\} }{(1-\eta_{A})^{2}} & \frac{\text{sin}^{2}\text{\ensuremath{\theta_{1}}}\text{sin}2\text{\ensuremath{\theta_{2}}}\exp(-\text{\ensuremath{i\phi_{2}})}}{2(1-\eta_{A})}\\
\frac{\sin2\text{\ensuremath{\theta_{1}}}\text{sin}2\theta_{2}\exp(i\phi_{12})}{4(1-\eta_{A})^{2}} & \frac{\text{sin}2\text{\ensuremath{\theta_{1}}}\text{sin}^{2}\text{\ensuremath{\theta_{2}}}\exp(\text{\ensuremath{i\phi_{1}})}}{2(1-\eta_{A})} & \frac{\text{sin}^{2}\text{\ensuremath{\theta_{1}}}\text{sin}2\text{\ensuremath{\theta_{2}}}\exp(\text{\ensuremath{i\phi_{2}})}}{2(1-\eta_{A})} & \text{sin}^{2}\text{\ensuremath{\theta_{1}}}\text{sin}^{2}\text{\ensuremath{\theta_{2}}}
\end{array}\right),\end{array}\label{eq:rhoAout}
\end{equation}
 and 
\begin{equation}
\begin{array}{lcl}
\rho_{P,{\rm out}} & = & N_{P}\left(\begin{array}{cccc}
P_{11}\text{cos}^{2}\text{\ensuremath{\theta_{1}}}\text{cos}^{2}\text{\ensuremath{\theta_{2}}} & 2\text{\ensuremath{\cos}}^{2}\text{\ensuremath{\theta_{1}}}\text{sin}2\text{\ensuremath{\theta_{2}}}\exp(-i\phi_{2}) & 2\text{sin}2\text{\ensuremath{\theta_{1}}}\text{\ensuremath{\cos}}^{2}\text{\ensuremath{\theta_{2}}}\exp(-i\phi_{1}) & \text{sin}2\text{\ensuremath{\theta_{1}}}\text{sin}2\text{\ensuremath{\theta_{2}}}\exp(-i\phi_{12})\\
2\text{\ensuremath{\cos}}^{2}\text{\ensuremath{\theta_{1}}}\text{sin}2\text{\ensuremath{\theta_{2}}}\exp(i\phi_{2}) & 4\text{cos}^{2}\ensuremath{\theta_{1}}\text{sin}^{2}\text{\ensuremath{\theta_{2}}} & \text{sin}2\text{\ensuremath{\theta_{1}}}\text{sin}2\text{\ensuremath{\theta_{2}}}\exp(-i\Delta\phi) & 2\text{sin}2\text{\ensuremath{\theta_{1}}}\text{sin}^{2}\text{\ensuremath{\theta_{2}}}\exp(-i\phi_{1})\\
2\text{sin}2\text{\ensuremath{\theta_{1}}}\text{\ensuremath{\cos}}^{2}\text{\ensuremath{\theta_{2}}}\exp(i\phi_{1}) & \text{sin}2\text{\ensuremath{\theta_{1}}}\sin2\text{\ensuremath{\theta_{2}}}\exp(i\Delta\phi) & 4\text{sin}^{2}\text{\ensuremath{\theta_{1}}}\text{\ensuremath{\cos}}^{2}\text{\ensuremath{\theta_{2}}} & 2\text{sin}^{2}\text{\ensuremath{\theta_{1}}}\text{sin}2\text{\ensuremath{\theta_{2}}}\exp(-i\phi_{2})\\
\text{sin}2\text{\ensuremath{\theta_{1}}}\text{sin}2\text{\ensuremath{\theta_{2}}}\exp(i\phi_{12}) & 2\sin2\text{\ensuremath{\theta_{1}}}\text{sin}^{2}\text{\ensuremath{\theta_{2}}}\exp(i\phi_{1}) & 2\sin^{2}\text{\ensuremath{\theta_{1}}}\sin2\text{\ensuremath{\theta_{2}}}\exp(i\phi_{2}) & P_{11}\text{sin}^{2}\text{\ensuremath{\theta_{1}}}\text{sin}^{2}\text{\ensuremath{\theta_{2}}}
\end{array}\right),\end{array}\label{eq:rhoPout}
\end{equation}
 where $\phi_{12}=\phi_{1}+\phi_{2},$ $\Delta\phi=(\phi_{1}-\phi_{2})$,
$P_{11}=\frac{4\left(1-2\eta_{P}+2\eta_{P}^{2}\right)^{2}}{(1-\eta_{P})^{4}},$
\[
N_{A}=\frac{4(1-\eta_{A})^{4}}{2\left[\left(2-4\eta_{A}+5\eta_{A}^{2}-4\eta_{A}^{3}+2\eta_{A}^{4}\right)+\eta_{A}\left(2-3\eta_{A}+2\eta_{A}^{2}\right)\left\{ \text{cos}2\text{\ensuremath{\theta_{1}}}+\text{cos}2\text{\ensuremath{\theta_{2}}}\right\} +\eta_{A}^{2}\text{cos}2\text{\ensuremath{\theta_{1}}}\text{cos}2\text{\ensuremath{\theta_{2}}}\right]},
\]
 and
\[
N_{P}=\frac{(1-\eta_{P})^{4}}{2\left[\left(2-8\eta_{P}+14\eta_{P}^{2}-12\eta_{P}^{3}+5\eta_{P}^{4}\right)+2\eta_{A}^{2}\left(2-4\eta_{P}+3\eta_{P}^{2}\right)\text{cos}2\text{\ensuremath{\theta_{1}}}\text{cos}2\text{\ensuremath{\theta_{2}}}\right]}.
\]
Using (\ref{eq:fidelity}) and (\ref{eq:rhoAout}) we obtain the fidelity
of the quantum state prepared using the proposed CBST scheme under
amplitude-damping noise as

\begin{equation}
\begin{array}{lcl}
F_{{\rm AD}} & = & \frac{1}{16\left(2-4\eta_{A}+5\eta_{A}^{2}-4\eta_{A}^{3}+2\eta_{A}^{4}+\eta_{A}^{2}\text{cos}2\text{\ensuremath{\theta_{1}}}\text{cos}2\text{\ensuremath{\theta_{2}}}+\eta_{A}\left(2-3\eta_{A}+2\eta_{A}^{2}\right)\left(\text{cos}2\text{\ensuremath{\theta_{1}}}+\text{cos}2\text{\ensuremath{\theta_{2}}}\right)\right)}\\
 & \times & \left[32-164\eta_{A}+57\eta_{A}^{2}-26\eta_{A}^{3}+10\eta_{A}^{4}+\eta_{A}\left(34-51\eta_{A}+30\eta_{A}^{2}\right)\left(\text{cos}2\text{\ensuremath{\theta_{1}}}+\text{cos}2\text{\ensuremath{\theta_{2}}}\right)\right.\\
 & + & \eta_{A}^{2}\left(3-2\eta_{A}+2\eta_{A}^{2}\right)\left(\text{cos}4\text{\ensuremath{\theta_{1}}}+\text{cos}4\text{\ensuremath{\theta_{2}}}\right)+4\eta_{A}^{3}\left(3-2\eta_{A}+2\eta_{A}^{2}\right)\left(\text{cos}2\text{\ensuremath{\theta_{1}}}\text{cos}4\text{\ensuremath{\theta_{2}}}+\text{cos}4\text{\ensuremath{\theta_{1}}}\text{cos}2\text{\ensuremath{\theta_{2}}}\right)\\
 & + & \left.16\eta_{A}^{2}\left(2-2\eta_{A}+\eta_{A}^{2}\right)\text{cos}2\text{\ensuremath{\theta_{1}}}\text{cos}2\text{\ensuremath{\theta_{2}}}+\eta_{A}^{2}\left(1-2\eta_{A}+2\eta_{A}^{2}\right)\text{cos}4\text{\ensuremath{\theta_{1}}}\text{cos}4\text{\ensuremath{\theta_{2}}}\right]
\end{array}\label{eq:fidelity-Amp-damp-probab}
\end{equation}
Similarly, by using (\ref{eq:fidelity}) and (\ref{eq:rhoPout}) we
obtain the fidelity of the quantum state prepared using the proposed
CBST scheme under phase-damping noise as

\begin{equation}
\begin{array}{lcl}
F_{{\rm PD}} & = & \frac{32-128\eta_{P}+210\eta_{P}^{2}-164\eta_{P}^{3}+59\eta_{P}^{4}+\eta_{P}^{2}\left\{ 2-4\eta_{P}+3\eta_{P}^{2}\right\} \left(16\text{cos}2\text{\ensuremath{\theta_{1}}}\text{cos}2\text{\ensuremath{\theta_{2}}}+\text{cos}4\text{\ensuremath{\theta_{1}}}\text{cos}4\text{\ensuremath{\theta_{2}}}+3\left(\text{cos}4\text{\ensuremath{\theta_{1}}}+\text{cos}4\text{\ensuremath{\theta_{2}}}\right)\right)}{16\left(2-8\eta_{P}+14\eta_{P}^{2}-12\eta_{P}^{3}+5\eta_{P}^{4}+\eta_{P}^{2}\left\{ 2-4\eta_{P}+3\eta_{P}^{2}\right\} \text{cos}2\text{\ensuremath{\theta_{1}}}\text{cos}2\text{\ensuremath{\theta_{2}}}\right)}.\end{array}\label{eq:fidelity-phase-damp-probab}
\end{equation}

Following a similar strategy, we can also obtain analytic expressions
for the fidelity for the CQD protocol described above in a noisy environment.
The analytic expressions of fidelity would depend on the Charlie's
choice of initial state and Alice's and Bob's secret messages (i.e.,
on unitary operators corresponding to their secret messages). Expressions
for fidelity obtained for various situations are listed in Table \ref{tab:AD-fidelity-CQD}. 

\begin{table}
\begin{centering}
\begin{tabular}{|c|c|c|>{\centering}p{4cm}|>{\centering}p{4cm}|}
\hline 
Initial state & Operations of Alice & Operations of Bob & Fidelity in amplitude damping channel $\left(F_{AD}^{\prime}\right)$ & Fidelity in phase damping channel $\left(F_{PD}^{\prime}\right)$\tabularnewline
\hline 
 & $X,\, iY$ & $X,\, iY$ & $F_{AD1}^{\prime}=\frac{4-8\eta_{A}+7\eta_{A}^{2}-2\eta_{A}^{3}+\eta_{A}^{4}}{4\left(1-\eta_{A}+\eta_{A}^{2}\right)}$ & \tabularnewline
\cline{2-4} 
$|\psi^{\pm}\rangle$  & $I,\, Z$ & $I,\, Z$ & $F_{AD2}^{\prime}=\frac{4-8\eta_{A}+9\eta_{A}^{2}-4\eta_{A}^{3}+\eta_{A}^{4}}{4\left(1-\eta_{A}+\eta_{A}^{2}\right)}$ & $F_{PD1}^{\prime}=\frac{2-6\eta_{A}+8\eta_{A}^{2}-4\eta_{A}^{3}+\eta_{A}^{4}}{2\left(1-2\eta_{P}+2\eta_{P}^{2}\right)}$\tabularnewline
\cline{2-4} 
 & $X,\, iY$ & $I,\, Z$ & $F_{AD3}^{\prime}=\frac{\left(1-\eta_{A}\right)^{2}\left(4+\eta_{A}^{2}\right)}{4\left(1-\eta_{A}+\eta_{A}^{2}\right)}$ & \tabularnewline
\cline{2-4} 
 & $I,\, Z$ & $X,\, iY$ & $F_{AD4}^{\prime}=\frac{4-8\eta_{A}+7\eta_{A}^{2}-4\eta_{A}^{3}+\eta_{A}^{4}}{4\left(1-\eta_{A}+\eta_{A}^{2}\right)}$ & \tabularnewline
\hline 
$|\phi^{\pm}\rangle$  & $I,\, X,\, iY,\, Z$ & $I,\, X,\, iY,\, Z$ & $F_{AD5}^{\prime}=\frac{1}{4}\left(2-\eta_{A}\right)^{2}$ & $F_{PD2}^{\prime}=\frac{1}{2}\left(2-2\eta_{P}+\eta_{P}^{2}\right)$\tabularnewline
\hline 
\end{tabular}
\par\end{centering}

\caption{\label{tab:AD-fidelity-CQD}Analytic expressions of fidelity for the
proposed CQD protocol in amplitude and phase damping channels for
all possible situations. Here\textcolor{red}{{} }the $\prime$ in superscript
is used to distinguish the fidelities of CQD protocol from that of
the BCST protocols. Here, we have considered that initially Charlie
prepared a Bell state $|\psi\rangle\in\{|\psi^{\pm}\rangle,|\phi^{\pm}\rangle\}$
and sent both the qubits to Alice. As both the qubits went through
the same channel, same Kraus operator worked on them. Later one of
the qubits travelled from Alice to Bob and Bob to Alice in a noisy
environment.}
\end{table}

\begin{figure}[H]
\begin{centering}
\includegraphics[angle=-90]{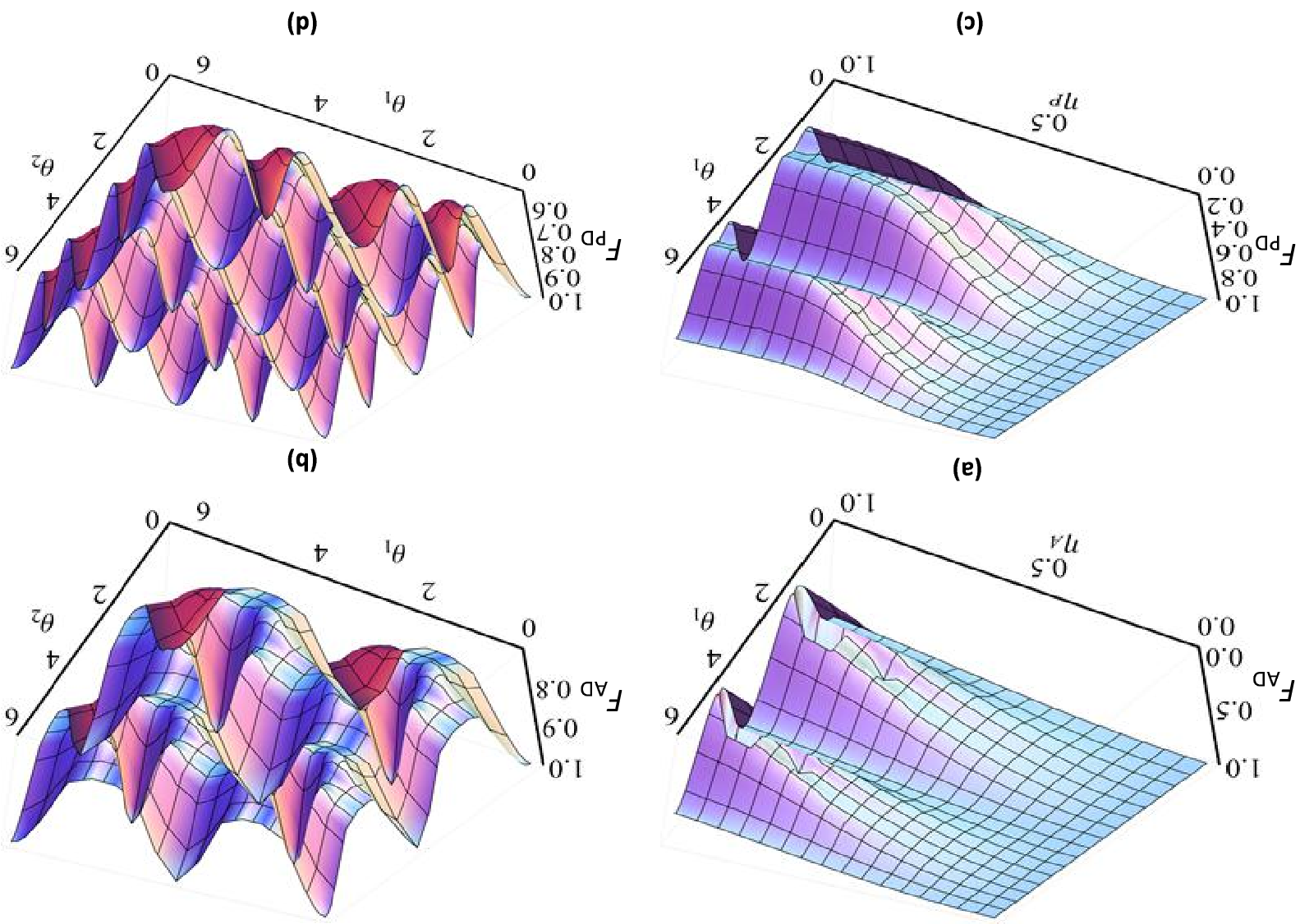} 
\par\end{centering}

\caption{\label{fig:ampdampand-phasedamp}(Color online) Comparison of the
effects of amplitude damping and phase damping noise on the fidelity
in Bell-state-based BCST protocol. (a) and (b) show the effect of
amplitude-damping noise, and (c) and (d) show the effect of phase-damping
noise on the fidelity of the Bell-state-base BCST protocol. (a) and
(c) show the effects of noise on the fidelity for $\theta_{2}=\frac{\pi}{6}$.
(b) and (d) show the effect of noise for $\eta=0.5$.}
\end{figure}

\begin{figure}[H]
\begin{centering}
\includegraphics[angle=-90,scale=0.5]{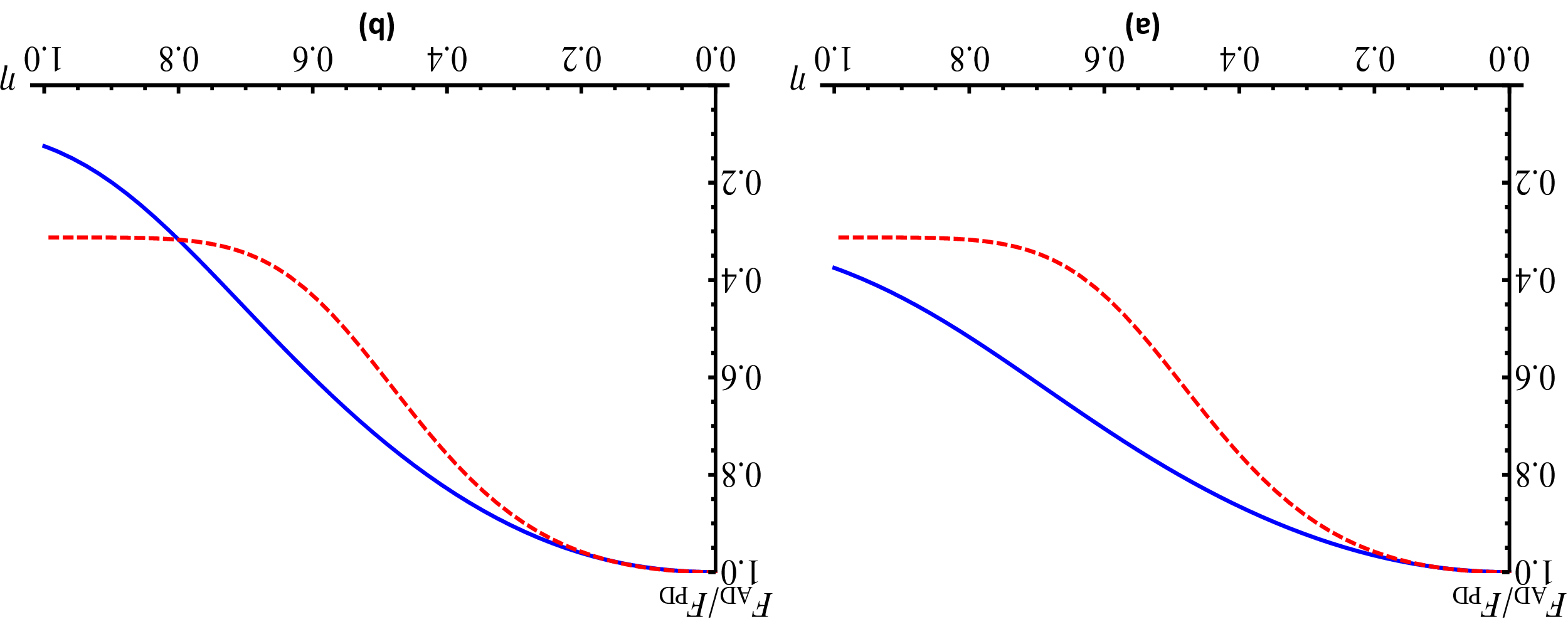}
\par\end{centering}

\caption{\label{fig:Comparison-of-ampdampand-phasedamp}(Color online) Comparison
of the effects of amplitude damping and phase damping noise on the
fidelity of Bell-state-based BCST scheme. The smooth (blue) and dashed
(red) lines correspond to the effects of amplitude-damping and phase-damping
channels on the fidelity of the Bell-state-based scheme of BCST, respectively.
The first plot shows the effect of noise on the fidelities for $\theta_{1}=\frac{\pi}{4}$
and $\theta_{2}=\frac{\pi}{6}$, and the second plot shows the effect
of noise on the fidelities for $\theta_{1}=\frac{\pi}{4}$ and $\theta_{2}=\frac{\pi}{3}$.}
\end{figure}

From (\ref{eq:fidelity-Amp-damp-probab})-(\ref{eq:fidelity-phase-damp-probab})
we can observe that the fidelities $F_{{\rm AD}}$ and $F_{{\rm PD}}$
are independent of the corresponding phase information $\phi_{i},$
and they depend only on the decoherence rate $\eta_{k}$ and the amplitude
information (i.e., $a_{i}$ and $b_{i}$) of the states to be teleported.
As the amplitude of the state used is fixed in the CQD protocol, fidelities
reported in Table \ref{tab:AD-fidelity-CQD} are naturally function
of $\eta_{k}$ only. These observations are consistent with the recent
observations reported in Ref. \cite{RSP-with-noise} in the context
of joint remote state preparation (JRSP) in noisy environment and
in Ref. \cite{CBRSP-our-paper} in context of controlled bidirectional
RSP (CBRSP). The method adopted to study the effect of noise in the
present paper and in Refs. \cite{RSP-with-noise,CBRSP-our-paper}
is quite general and can be easily applied to other schemes of quantum
communication and to study the effect of other noise models, such
as, generalized amplitude damping (GAD) channel or squeezed generalized
amplitude damping (SGAD) %
\footnote{AD and GAD channels are the special cases of SGAD channel. %
} channel \cite{switch,SGAD1}. Specifically, to study the effect of
SGAD or GAD channels on the fidelity, we will just require to replace
the Kraus operators used above by the Kraus operators of SGAD or GAD
channels, respectively. Similarly, one may study the effect of Pauli-type
noise, too \cite{pauli-type-noise}. However, here we restrict ourselves
to amplitude-damping and phase-damping channels only. In case the
above described Bell-state-based BCST scheme is realized in a noisy
environment, then the fidelity corresponding to various noise models
would depend on the various parameters as shown in Figs. \ref{fig:ampdampand-phasedamp}-\ref{fig:Comparison-of-ampdampand-phasedamp}.

\begin{figure}[H]
\begin{centering}
\includegraphics[angle=-90,scale=0.5]{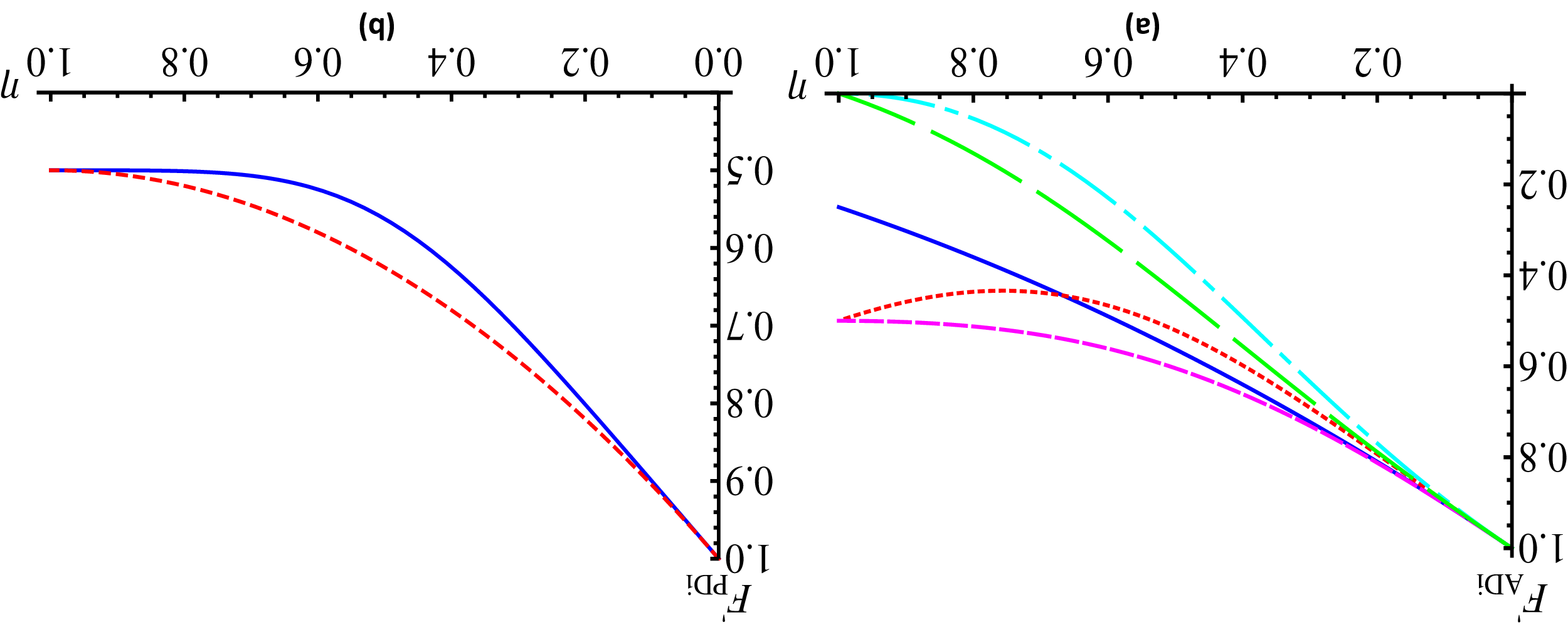}
\par\end{centering}

\caption{\label{fig:QD-Comparison-of-ampdampand-phasedamp-1}(Color online)
Comparison of the effects of amplitude damping and phase damping noise
on the fidelity for the protocol of CQD. (a) Compares all the fidelity
expressions for amplitude damping channel ($F_{AD1}-F_{AD5}$) provided
in the 4th column of Table \ref{tab:AD-fidelity-CQD}. Smooth (blue)
line corresponds to $F_{AD5}^{\prime}$, i.e., initial state $|\phi^{\pm}\rangle,$
and dotted (red), small dashed (magenta), dot-dashed (cyan) and large
dashed (green) lines correspond to $F_{AD1}^{\prime},\, F_{AD2}^{\prime},\, F_{AD3}^{\prime}$
and $F_{AD4}^{\prime},$ respectively (i.e., for various encoding
operations of Alice and Bob for the initial state as $|\psi^{\pm}\rangle$).
(b) The smooth (blue) and dashed (red) lines in the second plot correspond
to the effects of phase-damping channel for the CQD protocol realized
with the initial state $|\psi^{\pm}\rangle$ and $|\phi^{\pm}\rangle,$
respectively. Clearly, $|\phi^{\pm}\rangle$ provides a better choice
as it leads to higher fidelity.}
\end{figure}

Specifically, Fig. \ref{fig:Comparison-of-ampdampand-phasedamp} compares
the fidelity in amplitude damping and phase damping channels (solid
and dashed lines, respectively) in CBST for two independent choices
of unknown states to be teleported. In Fig. \ref{fig:Comparison-of-ampdampand-phasedamp}
a, we can easily observe that the amplitude damping channel gives
higher fidelity than the phase damping channel for all values of decoherence
rate $\eta_{A}=\eta_{P}=\eta$. However, for a different choice of
states to be teleported, we obtain that even phase damping channel
can have higher fidelity, especially at higher values of $\eta$ (cf.
Fig. \ref{fig:Comparison-of-ampdampand-phasedamp} b). Thus, we may
conclude that the amplitude damping channels do not have always fidelity
greater than that in the phase damping channels for the same value
of decoherence rate. The fact is further illustrated in Fig. \ref{fig:ampdampand-phasedamp},
where the variation in the fidelity is discussed in two cases: (1)
Considering that Bob teleports a state with fixed value of $\theta_{2}$
(i.e., fixed value of $a_{2}$ and $b_{2}$), which represents a quantum
state from a family of quantum states that differ only by the value
of the relative phase $\phi_{2}$. As fidelity is independent of $\phi_{2},$
all such states are equivalent for our purpose. Specifically, variation
in fidelity for this particular case is shown in Fig. \ref{fig:ampdampand-phasedamp}
(2) Considering that $\eta_{k}=\eta=0.5$ for both the directions
of communication (i.e., for Alice to Bob and Bob to Alice communication).
Here, it is important to note that form the symmetry in the expressions
of the fidelity in both types of noise channels, we can observe that
fidelity should remain unchanged if Alice and Bob interchange the
unknown state they want to teleport to each other using the same quantum
channel. Finally, the effects of amplitude and phase damping channels
on the fidelity for CQD protocol are illustrated in Fig. \ref{fig:QD-Comparison-of-ampdampand-phasedamp-1}.
From Fig. \ref{fig:QD-Comparison-of-ampdampand-phasedamp-1} b, it
can be easily seen that in presence of a phase damping channel, the
fidelity for the CQD protocol that uses $|\phi^{\pm}\rangle$ as initial
state is always greater than that which uses $|\psi^{\pm}\rangle$
as the initial state. Thus, for a phase damping channel, it would
be beneficial to start with $|\phi^{\pm}\rangle$as the initial state.
However, no such preference is observed in case of amplitude damping
channel (cf. Fig. \ref{fig:QD-Comparison-of-ampdampand-phasedamp-1}
a). Further, from Table \ref{tab:AD-fidelity-CQD}, we can observe
that the fidelities in the presence of phase damping noise are only
dependent on the initial state prepared by Charlie, while for the
amplitude damping channel, the fidelities are dependent on both the
initial state prepared by Charlie and the operations of Alice and
Bob.

\section{Conclusions \label{sec:Conclusions}}

We have already mentioned that in the recent past, several schemes
for controlled quantum communication (e.g., BCST, CQSDC, CQD, etc.)
have been proposed using $n$-qubit ($n\geq3$) entanglement. Specifically,
various protocols for BCST have been proposed using $m$-qubit entanglement
($m\in\{5,6,7\}$). Here, we propose a PoP-based protocol for BCST
that uses only Bell states and thus reduces the complexity of the
required quantum resources. We have also provided a Bell-state based
protocol of CQD and have studied the effect of amplitude damping and
phase damping channels on both the schemes. Further, we have shown
that a set of other schemes of controlled quantum communication (e.g.,
CQSDC, CQKD and CQKA) can be realized solely using Bell states. Extending
the discussion, here\textcolor{magenta}{{} }we note that it is straight
forward to realize that any channel that can be used for teleportation
can also be used for remote state preparation. Thus, the Bell-state-based
realization of BCST discussed above can be easily extended to provide
a protocol of CBRSP \cite{CBRSP-our-paper,ba-An-remote-state}. Interestingly,
in Refs. \cite{CBRSP-our-paper,ba-An-remote-state}, 5-qubit quantum
states are used to implement the same. In Ref. \cite{CBRSP-our-paper},
a scheme for controlled bidirectional JRSP (CBJRSP) was proposed using
the $7$-qubit quantum states of the form 
\begin{equation}
|\psi\rangle=\frac{1}{\sqrt{2}}\left(|GHZ_{1}\rangle_{123}|GHZ_{2}\rangle_{456}|a\rangle_{7}\pm|GHZ_{3}\rangle_{123}|GHZ_{4}\rangle_{456}|b\rangle_{7}\right)\label{eq:7-qubit-3}
\end{equation}
with $GHZ_{1}\neq GHZ_{3}$ and $GHZ_{2}\neq GHZ_{4}$ and $GHZ_{i}$
with $i\in\left\{ 1,2,3,4\right\} $ is a GHZ state. Following the
same logic as was adopted above to construct a Bell-state based protocol
of BCST, we can easily construct a GHZ-based protocol of CBJRSP. In
such a protocol, instead of 7-qubit state used in \cite{CBRSP-our-paper},
we will only require 3-qubit GHZ states. Clearly, PoP can be used
to perform various tasks of controlled quantum communication with
entangled states that involve a lesser number of qubits. As it is
easier to produce and maintain a Bell state or a GHZ state in comparison
to a multi-qubit entangled state, PoP essentially makes it easier
to implement the schemes of controlled quantum communication. Further,
the strategy used here to design the protocols of controlled quantum
communication has a few other advantages over the existing protocols
for the same task. For example, we may mention the following advantages:
(i) In this protocols, the controller can continuously vary the amount
of information revealed to the receiver(s) and the sender(s). (ii)
The controller has directional control in the BCST and similar protocols,
i.e., he can choose to disclose the information about the Bell state
required for the reconstruction of the teleported state in only one
direction (say the Bell state used for Alice to Bob teleportation)
without disclosing any information about the Bell state used for the
teleportation in the other direction (i.e., Bob to Alice teleportation).
Such control was not present in earlier schemes. Considering these
advantages, we conclude the paper with an expectation that the strategy
proposed here will play important role in the future development of
the protocols of controlled quantum communication.

\textbf{Acknowledgment:} AP and KT thank Department of Science and
Technology (DST), India for support provided through the DST project
No. SR/S2/LOP-0012/2010.

\end{document}